\documentclass[aip,rsi,floatfix,reprint]{revtex4-1}
\usepackage{hyperref} 
\hypersetup{
         pdfauthor={Ben Keitch; Vlad Negnevitsky; Weida Zhang},
         pdfborder={0 0 0},
         breaklinks=true}
\usepackage{siunitx}
\usepackage{graphicx}
\usepackage{subfig}

\makeatletter
\def\fps@figure{htbp}
\makeatother

\begin{document}

\bibliographystyle{aipnum4-1}

\title{Programmable and scalable radio frequency pulse sequence generator for multi-qubit quantum information experiments}

\begin{abstract}
  We present a versatile rf pulse control system that has been designed for multi-qubit quantum experiments.
  One instrument can be scaled to provide 32 channels of rf between 10 -- 450\,MHz.
  Synchronization can be achieved across multiple instruments.
  By using direct digital synthesis and custom control circuitry contained within a field-programmable gate array, sequences of transform-limited pulses can be produced.
  These have been used to carry out quantum gates that are able to meet fault-tolerant thresholds for single-- and two--qubit gate fidelities, as published elsewhere.
  We have also extended the frequency to the gigahertz regime using additional mixers to address hyperfine transitions in atomic systems.
  The system uses an efficient memory management scheme and a low-latency communications protocol that allows pulse sequences to be updated in real-time.
  Together these can enable outcome-based algorithms such as quantum error correction to be executed.
  The system is fully programmable in C++, and other languages such as Python can be supported by the on-board CPU, offering a highly flexible platform for a wide variety of experimental systems, and has been proven in trapped-ion quantum information experiments.
\end{abstract}

\author{Ben Keitch}
\altaffiliation{Present address: BAS, High Cross, Madingley Road, Cambridge CB3 0ET, UK}
\email{benitc@bas.ac.uk}
\affiliation{Institute for Quantum Electronics, ETH Z\"urich, Otto-Stern-Weg 1, 8093 Z\"urich, Switzerland}
\affiliation{Department of Engineering Science, University of Oxford, Parks Road, Oxford OX1 3PJ, UK}
\author{Vlad Negnevitsky}
\email{nvlad@phys.ethz.ch}
\affiliation{Institute for Quantum Electronics, ETH Z\"urich, Otto-Stern-Weg 1, 8093 Z\"urich, Switzerland}
\author{Weida Zhang}
\affiliation{Department of Engineering Science, University of Oxford, Parks Road, Oxford OX1 3PJ, UK}
\date{11 October 2017}
\maketitle

\section{Introduction}\label{sec:introduction}

Quantum information experiments generally encode information in the quantum states of an atom, molecule or solid-state system\,\cite{ladd_quantum_2010}.
A quantum system is prepared in a well-defined fiducial or superposition state, and quantum gates are performed by driving transitions between this and other states, usually using electromagnetic pulses.
In multilevel systems such as cold atoms and ions, further operations are used to detect the state population, and possibly to control excitation in motional modes, using such methods as Doppler and resolved-sideband cooling\,\cite{wineland_experimental_1998, barrett_sympathetic_2003}.
In this paper we will primarily discuss the application of a novel rf control system to trapped-ion experiments, in which the above operations are carried out using optical or microwave pulses.
Next we discuss the key system requirements and how they are met.

Atomic quantum systems require addressing of multiple transitions in addition to the qubit for full control.
For example, to carry out algorithms on two ions of different species, up to twenty laser beams may need to be controlled. Additionally, in physical systems there are often slow changes or drift in these resonant frequencies. To handle these demands complex control electronics and sophisticated software are required\,\cite{blatt_entangled_2008}.

The optical pulses used to address transitions are controlled with rf frequency pulses, via either acousto-optic (AO) or electro-optic (EO) modulators. 
AO and EO modulators are typically driven by rf at frequencies between 20\,MHz and several GHz, up to a typical power of 40\,dBm.
Pulses may also be applied directly in the rf or microwave domain using near-field antennas, over a broader frequency range including into the microwave band\,\cite{harty_high-fidelity_2016-1}.

Robust quantum algorithms carry significant overhead in both the number of physical systems needed for an error-tolerant qubit and the number of pulses per logical gate, and generally require ongoing classical feedback\,\cite{bermudez_assessing_2017}.
To achieve this the pulse sequencing system must have a large storage capacity, tens of channels, and low-latency decision-making and feedback.

In this paper, we describe a system developed around a backplane that supports modular rf channel cards and one master controller. Each card provides four rf channels and digital I/O, and 8 cards fit on one backplane, allowing for a total of 32 rf channels per rack-based instrument.
Most of the system hardware was designed in partnership with Enterpoint Ltd.\,\cite{enterpoint_ltd._milldown_2014} who now sell both the rf channel card and backplane under the name Milldown.

The rf is generated using the Direct Digital Synthesis (DDS) technique and the control logic is distributed between Field-Programmable Gate Arrays (FPGA) and general-purpose processors, allowing the system to be be programmed in a high-level language.
Each rf channel has an output range of 10 -- 450\,MHz, suitable for driving Zeeman and motional sidebands, via an AO modulator on a resonant atomic transition.
Driving hyperfine transitions at microwave frequencies has also successfully been carried out\,\cite{allcock_microfabricated_2013}
The frequency and phase of each rf pulse is \emph{deterministically} controlled  to a resolution of 233\,mHz (32-bit tuning word) and 96\,$\mu$rad (16-bit tuning word).
These parameters are repeatable and finely adjustable.
Pulse lengths are adjustable in 8\,ns increments, or 0.4\,\% for a 2\,$\mu$s pulse.
Pulse voltage is adjustable to 0.006\,\%, although it is subject to some thermal drift.
As fine calibration is always required to compensate for the non-linearities of external amplifiers and AO/EO modulators this does not contribute significantly to overall performance.

The firmware that was developed has been designed with both scalability and advanced algorithms in mind.
The overall pulse sequence is pre-programmed yet both the sequence and the individual pulse parameters (frequencies, phases, amplitudes and durations) can be updated dynamically as the sequence is running. This allows for deterministic feedback; this is discussed in more detail in section \ref{sec:system-overview}.

The combination of low-noise hardware and sophisticated firmware has already played a key role in novel trapped-ion experiments\,\cite{lo_spin-motion_2015, kienzler_observation_2016, leupold_sustained_2017, fluhmann_sequential_2017}.
With different firmware the rf cards have also been used in high-fidelity gate operations\,\cite{harty_high-fidelity_2014, ballance_high-fidelity_2016}.

The features of the rf system would also benefit a wide variety of closely-related quantum platforms including cold atoms and molecules, NMR and various solid-systems, and may be useful in broader applications such as spectroscopy, precision quantum measurement, or quantum chemistry.
The system could also be used in engineering applications, such as phased-array sonar or radar, or automated testing.

\section{RF Generation}\label{sec:rf-generation}

DDS is a reliable and coherent technique for rf generation that is already well-established and has met the stringent demands of quantum information experiments\,\cite{langer_high_2006, schindler_quantum_2013}.

A DDS consists of a Numerically Controlled Oscillator (NCO) connected to a Digital to Analogue Converter (DAC).
The phase register of the NCO can be changed arbitrarily, along with the step size, allowing full control of the phase and frequency of the rf.

In comparison to a Phase-Locked Loop (PLL), a major advantage of a DDS system is that its output frequency and phase can be deterministically changed with a far lower latency.
DDS also offers much finer frequency and phase resolution, as well as rapid ``hopping'' of parameters; i.e. they are frequency agile.

Another common approach is an Arbitrary Waveform Generator (AWG), which uses digital circuitry such as an FPGA followed by a DAC to output an arbitrary waveform rather than only a sine wave.
This is more flexible, and also offers two-tone or broadband outputs.
However, the downsides of an AWG of comparable performance to a DDS are greater complexity in the digital domain at high frequencies, higher power consumption, and more complex electronic hardware design and verification to achieve comparable noise levels.
We show that all the features we require can be implemented using DDS combined with other circuitry while giving the phase-noise performance quantum systems demand.

\section{System Overview}\label{sec:system-overview}

\begin{figure*}
\centering
\includegraphics[width=0.6\textwidth]{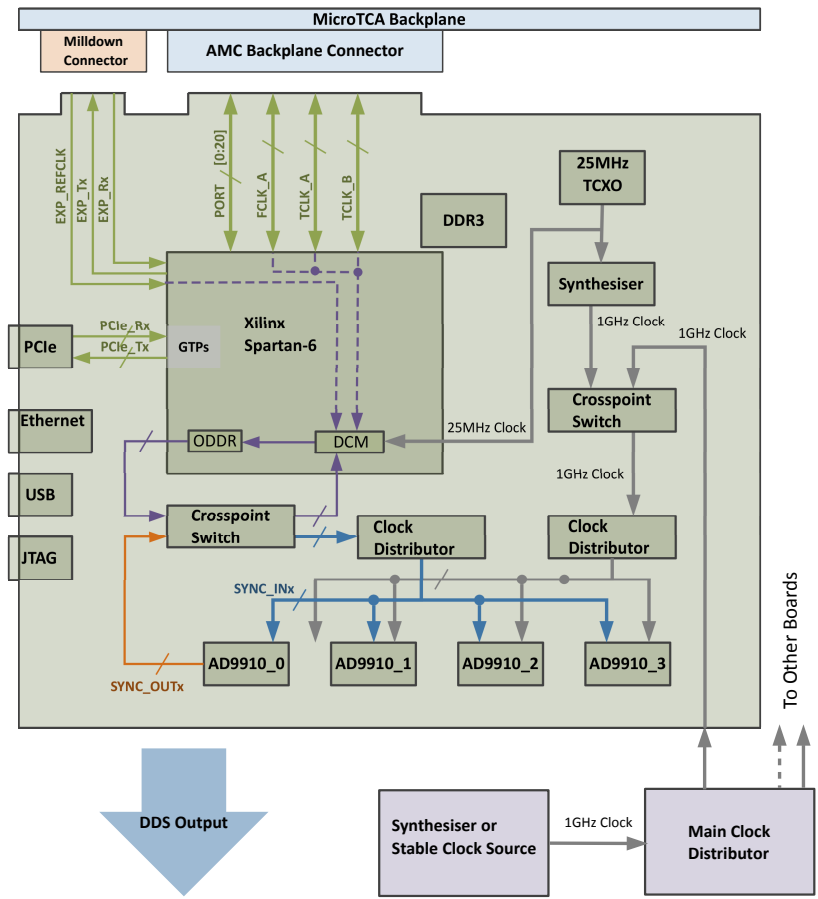}
\caption{Block diagram of a channel card showing backplane connections and clock distribution.}
\label{fig:block_diagram}
\end{figure*}
\begin{figure*}
\centering
\includegraphics[width=0.6\textwidth]{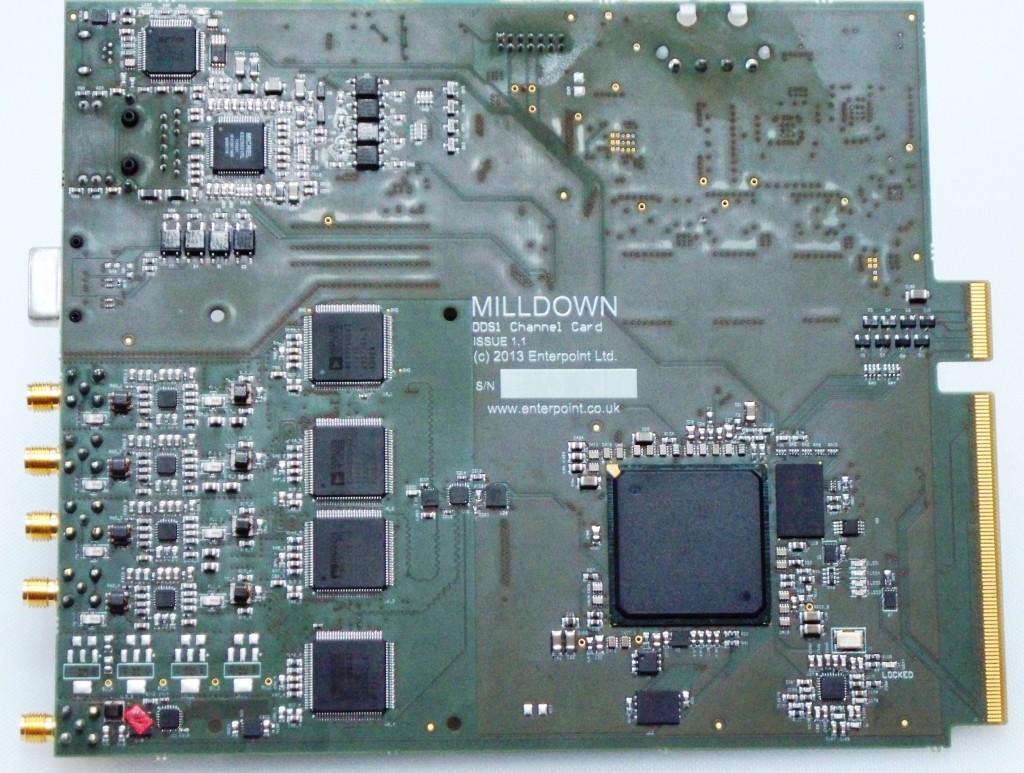}
\caption{Photograph of channel card showing the finger connection and SMA outputs.}
\label{fig:channel_card}
\end{figure*}

A block diagram of a single rf channel card is shown in figure \ref{fig:block_diagram}.
The channel card supports USB, Ethernet, backplane, and PCIe communications.
A photograph is shown in figure \ref{fig:channel_card}.
The channel cards are either inserted in a backplane or can be used standalone with a suitable power supply.
Being modular, the number of rf channels can be chosen from 4 to 32 per instrument. These are all phase-synchronous to one master clock.
The reference oscillator used determines the level of phase noise on the output channel, as we show below.
We achieve the required phase noise using a commercial integrated circuit and the correct choice of reconstruction filter.

The Analog Devices AD9910 DDS chip was chosen as it meets requirements for spurious digital harmonics and phase noise. It has 32-bit frequency tuning, 14-bit amplitude and 16-bit phase resolution. Frequency, phase and amplitude are set within 1.4\,$\mu$s via an SPI bus or more rapidly via a parallel bus\,\cite{analog_devices_ad9910_2016}. In the current firmware only the phase is updated using the parallel bus.
The fundamental output frequency spans 40 octaves from \SI{10}{\micro \hertz} to \SI{450}{\mega \hertz}, though the analog rf chain passes a minimum frequency of 10\,MHz. This covers all the required frequencies for trapped ions, except for direct addressing of the hyperfine transitions (typically \SIrange{1}{12}{\giga\hertz}). To reach hyperfine frequencies standard single-sideband generation methods were employed.
DDS also allows for near-perfect, exactly repeatable synchronization of multiple channels, allowing quadrature and other phase offset relationships to be easily accomplished\,\cite{analog_devices_fundamentals_2009, shappert_spatially_2013}.
Digital control logic, including timing control and internal memory for the four channels of rf on each channel card is provided by a Xilinx Spartan 6 FPGA. We use this to distribute the real-time sequencing tasks between the DDS channels. If multiple channel cards are used on a backplane system then a master card can be used to provide overall system control; this is the approach used at ETH and Oxford.
Multiple instruments could use Ethernet and a shared clock to scale beyond 32 channels.

\subsection{Backplane Architecture}\label{sec:backplane}

The Micro Telecoms Adaptor (mTCA) standard\,\cite{picmg_consortium_microtca_2017} was chosen as the backplane architecture.
This is a telecommnications and HPC standard, an extended version of which (mTCA.4) is used in high-energy physics experiments.
mTCA has low-latency interconnects and is an open standard, which makes it very attractive for the scientific community. 

The channel cards fit into a 19 inch sub-rack called a ``shelf'' in TCA terminology, which also contains an active backplane with finger-edge connectors.
A custom finger-edge connector was added to the channel card for supplying linearly regulated power and accessing Gigabit transceivers.
A custom passive backplane was designed to accommodate this.
We chose to adapt a standard Eurocard rack to mount this in, and to not implement the extra power and thermal management required by the mTCA spec.
Standard mTCA cards can fit the rack, but the channel cards will not fit a standard mTCA backplane without modification.
The custom backplane can support one master card (MCH) and up to eight channel cards or other mTCA cards such as commercial DAC cards.

A commercial MCH could be used, but we chose to implement a low-cost solution consisting of a Zedboard from Avnet \cite{avnet_zedboard_2017} attached to a custom adapter card that routes communication signals to the backplane via its FMC connector.
This uses a Xilinx Zynq chip consisting of an FPGA and dual-core ARM Cortex CPU, with access to 4Gb of off-chip DRAM.
It has hardware support for the Ethernet and USB physical layers.
The Zedboard is well-supported in the scientific and hobbyist communities.
FPGA firmware on the Zedboard and the channel cards provides real-time synchronization.
A software GUI running on the PC is used to set experimental parameters, and to process and display results.
Ethernet is used to communicate between the PC and Zedboard, and debugging is done over USB.

Quantum state detection for our experiments is based on a threshold decision made by counting photons emitted from the ion.
The FPGA can receive digital inputs (TTL level) from up to eight photon counters, which is sufficient for experiments with tens of ions.
Measurement-based outcomes are interpreted by the algorithm running on the ARM, which can then alter the running sequence on the channel cards in real-time.
To allow the system to scale, it is necessary to avoid a bottleneck at the MCH.
Pulse sequencing tasks are distributed amongst the channel cards and thus intra-board communication is kept to a minimum.
Quantum error syndrome measurement and correction could be performed within 40\,$\mu$s using these features.

\begin{figure*}
	\centering
	\includegraphics[width=0.6\textwidth]{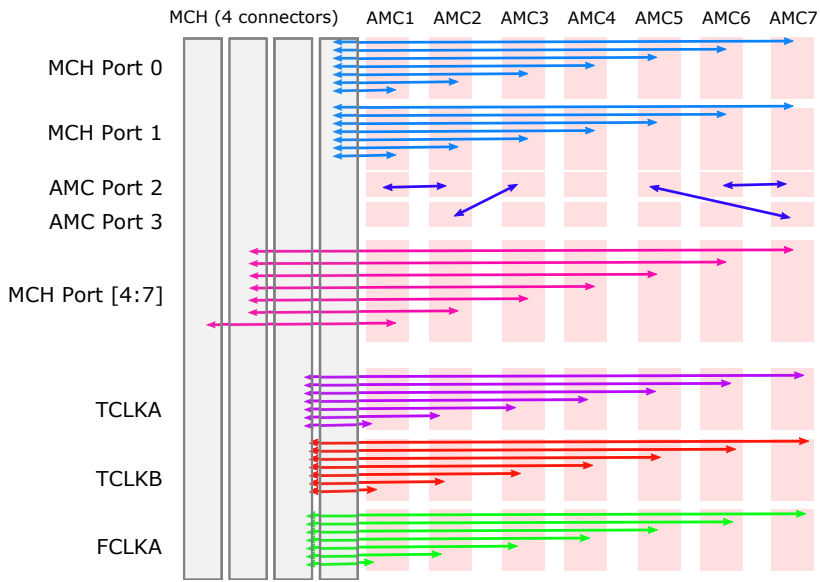}
	\caption{Simplified overview of how the backplane is connected in a single MCH MicroTCA system.}
	\label{fig:microtca}
\end{figure*}

As shown in figure \ref{fig:microtca}, the backplane provides impedance-controlled differential pair lanes arranged in both point-to-point and bus topologies.
Electrically, these can support speeds of up to 40\,Gb/s.
Geographical addressing allows the MCH to talk to one board at a time, or to send a broadcast command to all the boards simultaneously.
The MCH also generates trigger pulses transmitted over a shared clock lane of the backplane to keep all the channel cards synchronized. 

The point-to-point connections provide direct links without switching, which avoids unpredictable latencies and jitter which would be unacceptable in a quantum information system.
PCIe or Ethernet could be used but would add unnecessary complexity and latency; to use the full speed of the bus, gigabit transceivers would be required at both ends.
Instead, we use a simpler differential serial communication protocol to reach speeds sufficient for foreseeable trapped-ion experiments.
No specialist physical adaptor chip is required, as the FPGAs are directly capable of differential communication at these speeds.
With this the MCH can communicate reliably with the rf cards at up to 166\,Mb/s without encoding.
The protocol uses a 48-bit combined address/instruction word, which is written directly into the channel card memory with a latency of several clock cycles, below 20\,ns.
This allows us to change the pulse sequences and parameters in real-time with deterministic timing.

\section{Channel Card Details}\label{sec:channel-card-details}
\subsection{Hardware}\label{sec:hardware}

Each channel card has one FPGA and four channels of rf, consisting of a DDS chip, variable-gain amplifer (VGA) and control DAC, and filtering.
The Spartan 6 FPGA provides 4.8\,Mb of BRAM internal memory and 540 I/O pins that implement the various communications channels mentioned above as well as separate parallel buses to each DDS chip.

Finite-length rectangular pulses result in a sinc-like excitation profile in Fourier space centred on the pulse frequency, due to the step change in power at the start and end of the pulse.
This causes undesired off-resonant excitation of nearby transitions, however it can be reduced by smoothly \emph{shaping} the turn-on and turn-off power curve of the pulse to avoid rapid changes.
This helped increase multi-qubit gate fidelity by over 10\,\% in early experiments\,\,\cite{riebe_process_2006}.

\begin{figure*}
	\centering
	\includegraphics[width=0.8\textwidth]{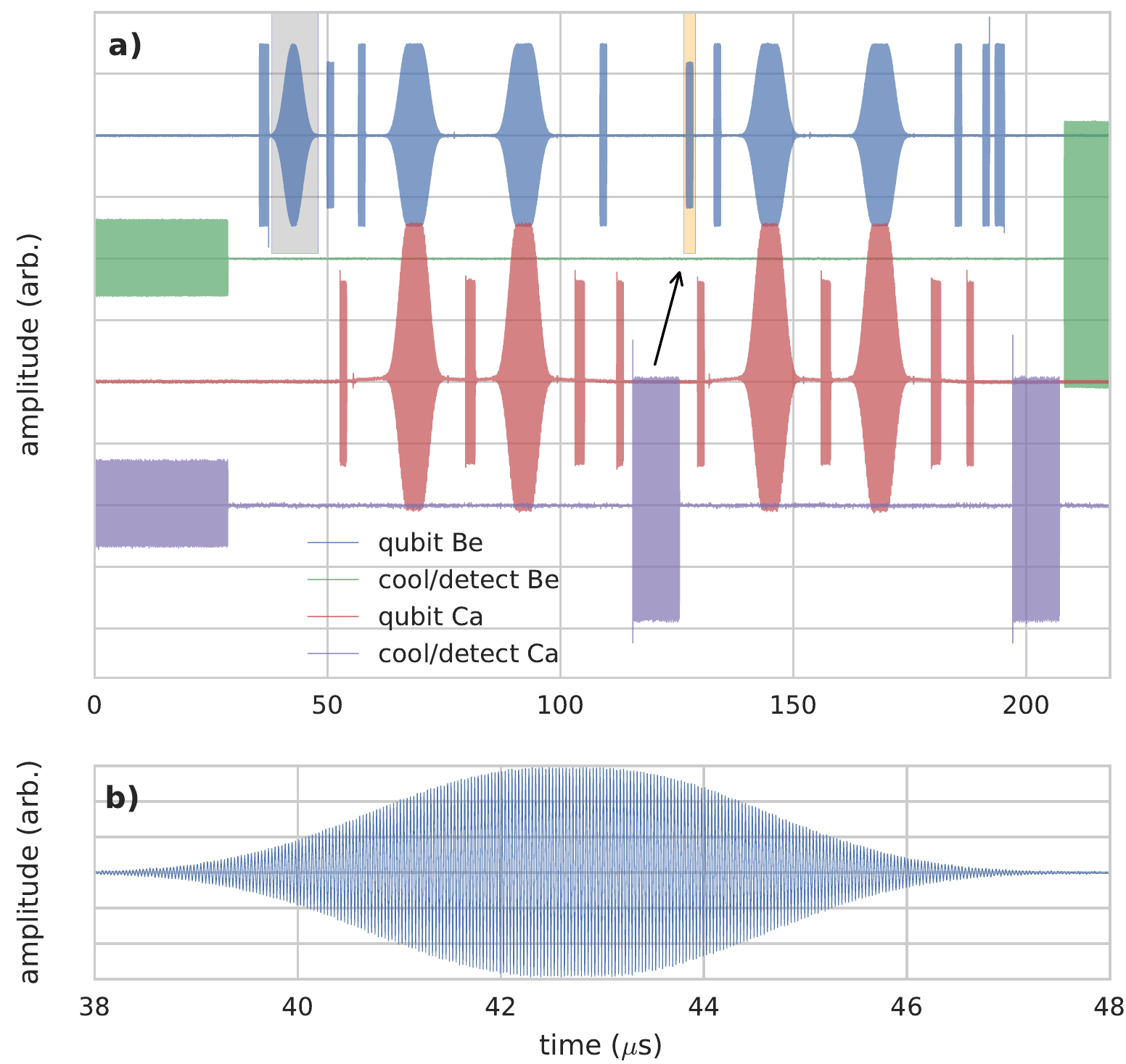}
	\caption{\textbf{a)} Oscilloscope traces of the DDS outputs for a typical mixed-species Be$^+$--Ca$^+$ pulse sequence, with the arrow indicating a pulse executed conditionally on the detection outcome. \textbf{b)} Expanded view of grey box in \textbf{a)}, showing a typical shaped pulse produced by the DDS.}\label{fig:pulse_seq}
\end{figure*}

To achieve the desired windowing functions, a dynamic range of \SIrange{30}{40}{\decibel} on the rf amplitude is required.
Using the amplitude register of the AD9910 DDS for both pulse shaping and overall amplitude control causes quantisation issues at lower power, because only a small part of the dynamic range of the internal DAC is utilised.
As a result, the pulse spectrum acquires power-dependent artefacts that are difficult to compensate.
To this end each rf channel includes an external variable-gain amplifier (VGA).
The DDS amplitude is still used for setting coarse amplitude.
A typical example of a multichannel sequence using shaped pulses is shown in figure \ref{fig:pulse_seq}.

The VGA chosen has a \SI{3}{\mega \hertz} bandwidth and a logarithmic response.
This gives a greater dynamic range than a linear VGA, with the drawback that the shaping waveform must be pre-compensated to linearise the response, adding a little extra complexity.

The VGA is controlled via a 14-bit 275\,MSPS DAC (Texas Instruments DAC5672A) which we run at \SI{62.5}{\mega \hertz} bandwidth.
This DAC has a higher bandwidth than the VGA, allowing us to overdrive the VGA at below-unity gain.

Each output is filtered to remove alias frequencies, and then further amplified to around 0\,dBm over the range \SIrange{10}{450}{\mega \hertz}.
To drive an AOM we amplify this to a few watts using an external amplifier whereas a small EOM can be driven directly.

The DDS chips are clocked from a single 1\,GHz source.
This clock signal can either be provided externally via a front panel SMA connector, or generated on-board for testing.
Clock signals can also be routed via the backplane (see figure \ref{fig:block_diagram}), however these must pass through the FPGA to reach the DDSes, which may be non-ideal for some applications.
We use a stable oscillator slaved to a rubidium clock and distributed to each card to ensure stable phase.
We chose not to discipline the rubidium clock from GPS as we can use our trapped ion as a frequency reference and adjust phases to the ion's atomic resonance.
The DDS clock is divided by 16 in the AD9910 to generate a signal used to synchronize all output channels.

\subsection{Firmware}

The MCH globally triggers every rf channel via the backplane.
Each rf channel runs a separate pulse sequence, which is encoded as a set of predefined pulse events; each event occurs a deterministic time (number of clock cycles) after the previous.
One common event is an edge, which has a defined wait time and an output frequency/phase/amplitude. The edge amplitude can also be shaped to achieve the transform-limited pulses discussed above.

This architecture minimises memory requirements for pulse sequences made up of arbitrarily-ordered but repetitive pulse events.
A typical sequence can be thousands of pulses long, yet its description needs less than 2\,kB of memory in the FPGA, as the repetitive structure can be efficiently exploited.
The channel card has off-chip memory available, but this has not yet been needed.
This RAM could in theory store millions of unique pulses.

Synchronising the phase of the rf with the event timing requires careful FPGA firmware design.
For each rf edge, the DDS phase is set to the required phase (DDS frequency multiplied by total pulse sequence time) via the parallel bus.
This mechanism ensures that the rf output is phase-coherent with respect to the reference frame of the quantum system being driven, even if multiple pulses at other frequencies have been run since the first pulse.

\section{System Performance}

A key design goal was to ensure that the electronic noise from the rf sources would not limit the fidelity of quantum experiments using the system.
Several noise types were investigated, and their sources are discussed in this section with this goal in mind.

\subsection{Power Supply Design}\label{sec:power-supplies}

Power supply noise is one of the most significant contributors to the DDS phase noise.
The rf phase noise contributes to decoherence and off-resonant carrier excitation of the ion, however this is not a dominant effect.
More significantly, the rf card power requirements place a large demand on the DC power source, with a tradeoff between efficiency and noise. Two approaches were investigated, standard linear power supplies and switch mode power (SMP) supplies.
SMP supplies typically work by switching current on and off into a load, and keep the load voltage constant by varying the duty cycle.
They operate at switching speeds around 50--500\,kHz.
This is more efficient than linear regulation, however it generates both injected noise on the supply rails, and radiated electromagnetic interference (EMI).
The former can be avoided with filtering, but EMI is more difficult to protect against, requiring good PCB design, physical distance, and in some cases metal cans for shielding.

Each channel card requires around 15\,W of power during normal operation, supplied at 12\,V and 3.3\,V.
Other voltages are generated with local regulators.
To evaluate the effect of power supply regulation on phase noise, two versions of the channel card were manufactured, a 'linear' card (Milldown Issue 1.1) with on-board linear regulators, and a 'switch-mode' card (Milldown Issue 1.2) with several switch-mode regulators for efficiency at high currents.
Three factors help suppress the power supply noise.
Each voltage used directly by the analog circuitry in the DDSes, DACs and rf chain is produced by low-noise linear regulators, while the switch-mode regulators only supply the linear regulator inputs and other voltages whose noise is less critical.
The Power Supply Rejection Ratio (PSRR) of the DDSes and DACs reduces supply noise effects inside the chips.
Common-mode output rejection is also effective, and is provided by the differential-current output structure plus the use of a balanced-unbalanced transformer (balun) at the rf output.
This is effective as a large portion of the power supply noise is common-mode.

An ATX power supply similar to those used in desktop PCs was chosen as the global power supply, providing up to 83\,A at 12\,V.
ATX supplies are normally switch-mode.
The quality of their filtering varies a lot, so care was taken to choose a high-quality ATX supply (Seasonic SS-1000XP). However, for comparison, measurements were also done using a linear bench supply.

Thermal fluctuations can cause rf output power instability, unless properly managed.
Forced-air cooling of the system is required, as both the linear and switch-mode cards reach about 80\,$^o$C in normal use.
A fan tray was mounted above and below the cards in a push-pull configuration and cooling fins were added to the FPGA and DDS chips.
This combination kept the board temperature and the rf power fluctuations within our requirements.
Electrical noise (EMI) from the fans was observed.
This could be reduced using a brushless type, or by ducting the air from a distance, however careful grounding reduced it to an acceptable level in the test setup.

\subsection{Phase noise}\label{sec:phase-noise}

\begin{figure*}
	\centering
	\subfloat[Issue 1.1 card with linear regulators]{
		\includegraphics[width=0.8\textwidth]{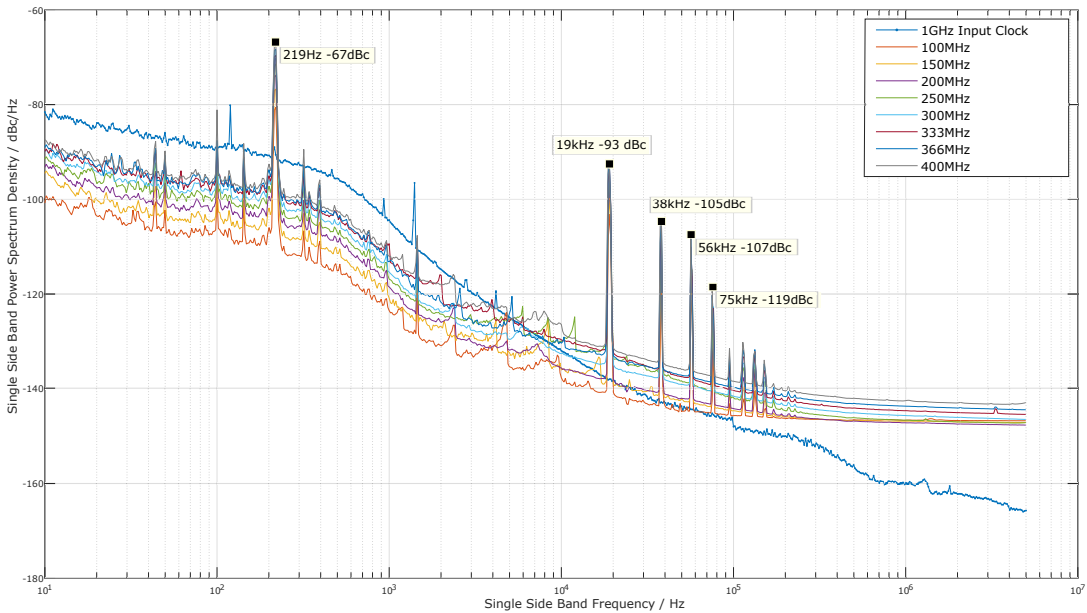}
	}\par
	

	\subfloat[Issue 1.2 card with switch-mode regulators]{
		\includegraphics[width=0.8\textwidth]{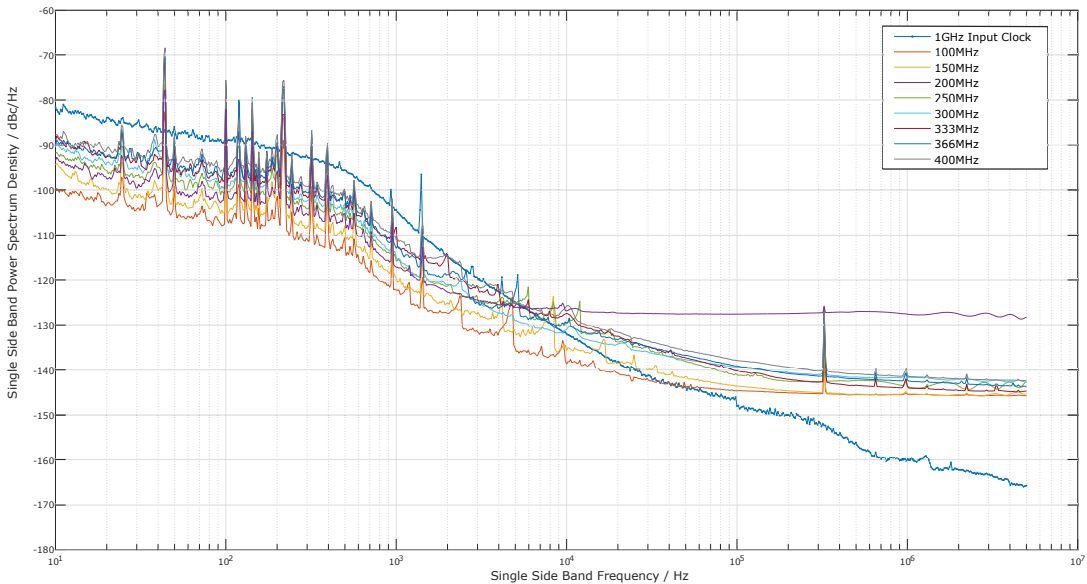}
	}
	\caption{Noise power spectral density for the two different cards.
          Different DDS output frequencies are shown alongside the free-running 1\,GHz clock for comparison.
          Spurs at 219\,Hz and harmonics of 19\,kHz have been labelled in (a)}
	\label{fig:phase_noise}
\end{figure*}

\begin{figure*}
  \centering
  \includegraphics[width=0.8\textwidth]{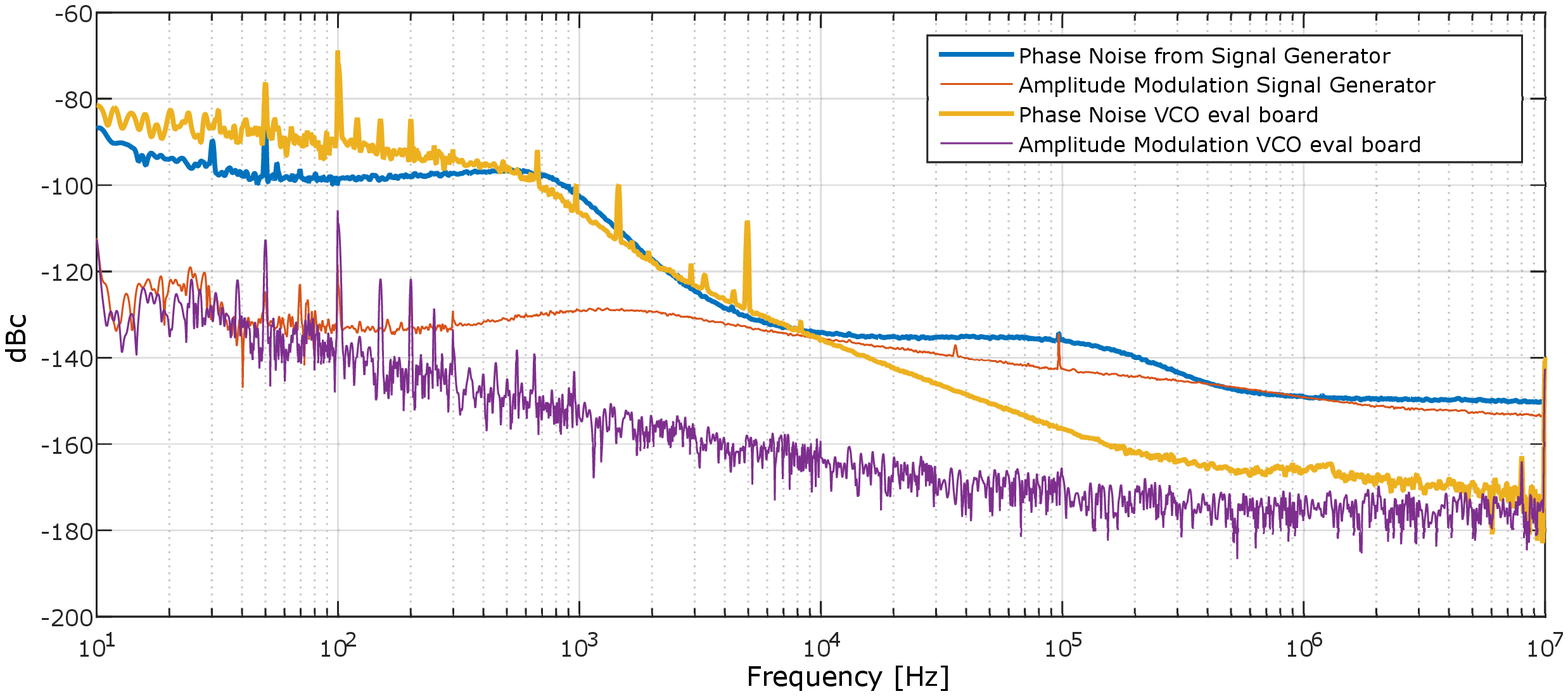}
  \caption{Phase noise and amplitude modulation noise of Agilent E4437B synthesiser vs an Analog Devices ADF4106 PLL evaluation board with a CVCO55CX1000 VCO, both running at 1\,GHz.}
  \label{fig:vco_noise}
\end{figure*}

The phase noise of the linear and the switch-mode cards has been measured under various conditions using a Keysight E5052B phase noise analyser.
A 1\,GHz clock signal for driving the cards was generated with a PLL evaluation board from Analog Devices (ADF4106) and a VCO (Crystek CVCO55CX-1000-1000).
This VCO has lower phase noise than high-quality instrument synthesizers, which must compromise phase noise for tunability.
The PLL was locked to the 10\,MHz reference output of the E5052B phase-noise analyser during measurements.
The results are shown in figure \ref{fig:phase_noise}.
The phase noise is comparable with the free-running VCO.
The DDS channel provides better noise performance than commercial frequency synthesisers as can be seen in \ref{fig:vco_noise}

\subsection{Unregulated spurs}\label{sec:unregulated-spurs}

The phase noise of the switch-mode card is better than the linear card in terms of Spurious-Free Dynamic Range (SFDR).
The spurs visible in the \SIrange{10}{200}{\kilo \hertz} band on the output of the linear card are related to the ATX power supply.
It was noted that changing the load current of the card causes the spurs to relocate on the spectrum by about 1\,kHz.

Replacing the ATX supply with a linear power supply lowered the overall noise floor by about 2\,dB, but did not alter the spurs.
It was found that the spurs were due to clock jitter not being rejected sufficiently by the on-board regulation.
Extra or better decoupling capacitors could solve this issue.

Although the overall SFDR of the switch-mode card is better than the linear card, the broadband noise floor is worse on the switch-mode card at frequencies above 1\,MHz by about 1\,dB.
This results in a large difference when considering the integral noise.
This is to be expected because the noise spectral density of the linear regulators is better than the switch-mode regulators; however, the switch-mode card is sufficiently good for our needs, and eases the demands on thermal management.
As it does not suffer from spurs as described above, these cards are preferred.

\subsection{Clock Jitter}\label{sec:clock-jitter}

\begin{figure*}
  \centering
  \includegraphics[width=0.8\textwidth]{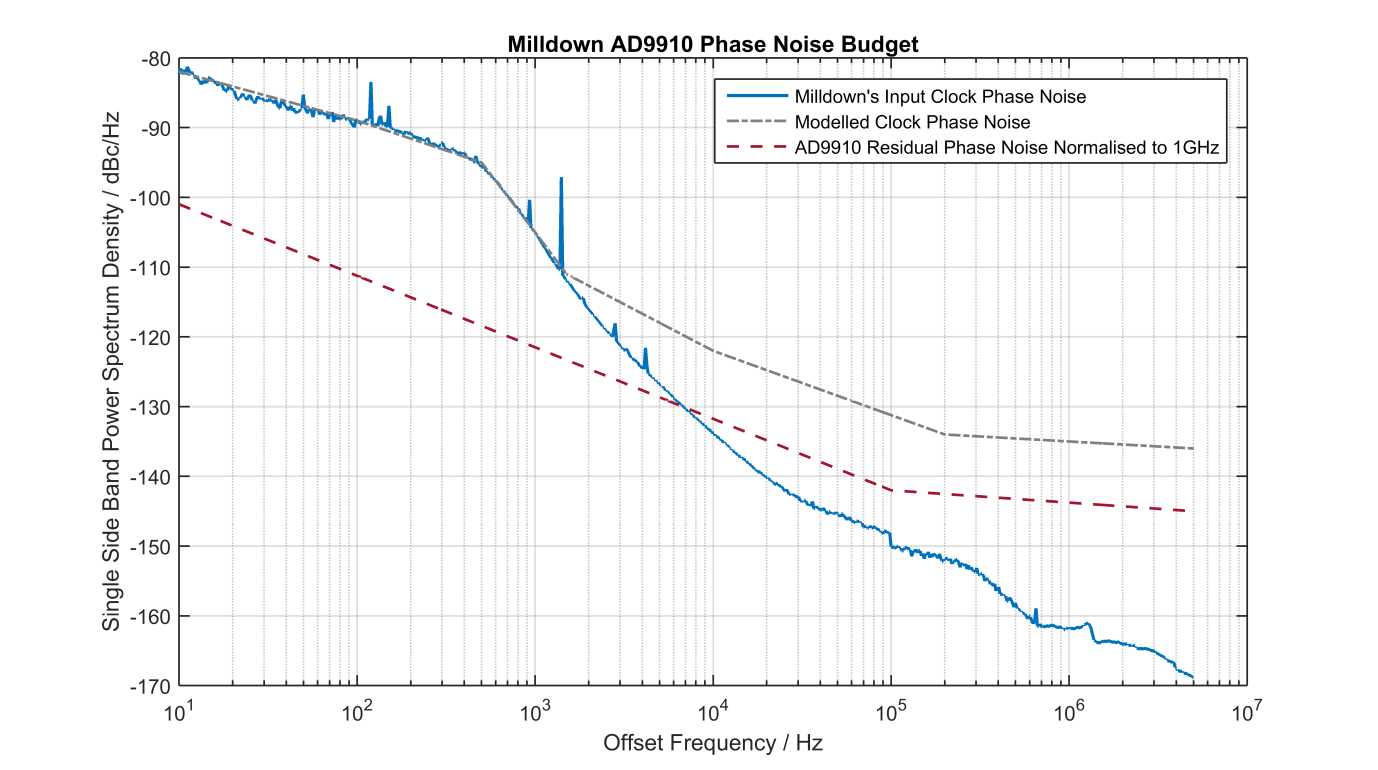}
  \caption{Residual noise of the DDS output after the input clock phase noise has been considered.}
  \label{fig:dds_residual}
\end{figure*}

Ideally the DDS output phase noise should follow the phase noise curve of the input clock, until a point where the residual phase noise of the DDS dominates, as shown in figure \ref{fig:dds_residual}.
The red line is the residual phase noise at 98.6\,MHz from the AD9910 datasheet, normalised to 1\,GHz.
The residual phase noise of the DDS is measured by removing those components in the DDS output phase noise which are cross-correlated with the input clock of the DDS and its power supply.

\begin{figure*}
  \centering \includegraphics[width=0.8\textwidth]{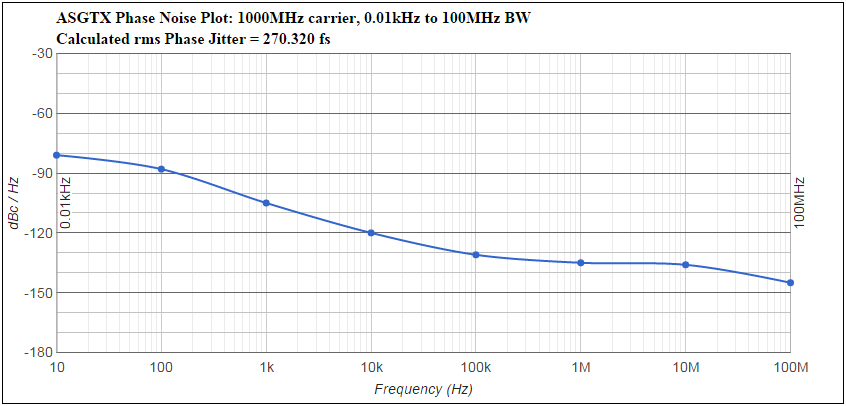}
  \caption{Plot of phase-jitter as modelled for the clock and distribution network.}
  \label{fig:jitter_plot}
\end{figure*}

To understand where the excess noise comes from, we modelled a 1\,GHz clock at the floor of the normalised DDS output phase noise, as shown in figure \ref{fig:jitter_plot}.
The normalised phase noise converges to this modelled clock, which is notably higher than both the input clock phase noise and the residual phase noise of the AD9910.
This implies the actual clock input of the DDS is worse than the clock input of the board, due to the clock distribution network between the input clock and the DDS input, which consists of a fanout chip and crosspoint switch as shown in figure \ref{fig:block_diagram}.
We calculate the RMS jitter of the modelled clock in the range 10\,Hz to 100\,MHz to be 270\,fs.
Comparing this to the manufacturer's specified values for the crosspoint switch and the clock fanout, we see this is within the values given of 500\,fs and 86\,fs respectively.

Therefore, phase noise increase between the board input clock and the DDS output (in the 10\,kHz to 5\,MHz offset) is most likely due to the additive jitter of the 1\,GHz clock distributing network on the rf card.
This can be avoided using direct passive connections to the DDS chips. Future versions will include an SMA connector for this purpose.

\subsection{Noise Spectral Density of the VGA}
\label{sec:noise-spectral-density}

\begin{figure}
\centering
\includegraphics[width=\columnwidth]{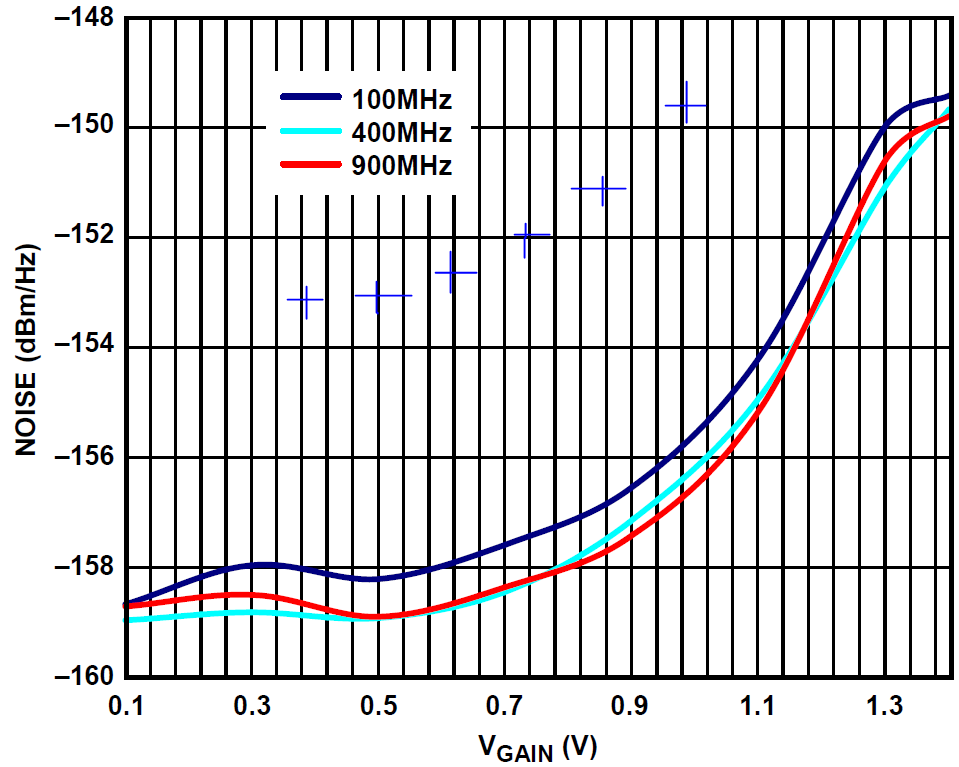}
\caption{Plot of actual noise power density (blue crosses) and specified noise (solid lines) due to the Variable Gain Amplifier.}
\label{fig:vga-plot}
\end{figure}

The datasheet of the VGA (Analog Devices ADL5330) shows typical output noise spectral density vs the gain controlling signal (V\textsubscript{gain}) reproduced here in figure \ref{fig:vga-plot}. The output noise spectral density is plotted at 100, 200 and 400\,MHz offset from the carrier.

We measured noise from the DDS card at 5\,MHz offsets, whilst controlling the VGA gain from 0\,dB to -15\,dB. These values of spectral noise density vs VGA control voltage are plotted, overlaid as blue crosses on the datasheet graph in figure \ref{fig:vga-plot}.

The measured noise follows the same trend as the VGA's specified output noise for relative attenuation, but about 5\,dB higher.
This additional 5\,dB of noise is due to the signal source used not being as noise-free as in the manufacturer's test.
The plotted results suggest that the VGA dominates the broadband noise (density) floor, as it is the final component on the rf chain.
Future card versions will provide a bypass for the VGA when its use is not absolutely necessary.

\section{Frequency Upconversion}\label{sec:frequency_upconversion}

\begin{figure*}
	\centering
	\includegraphics[width=0.8\textwidth]{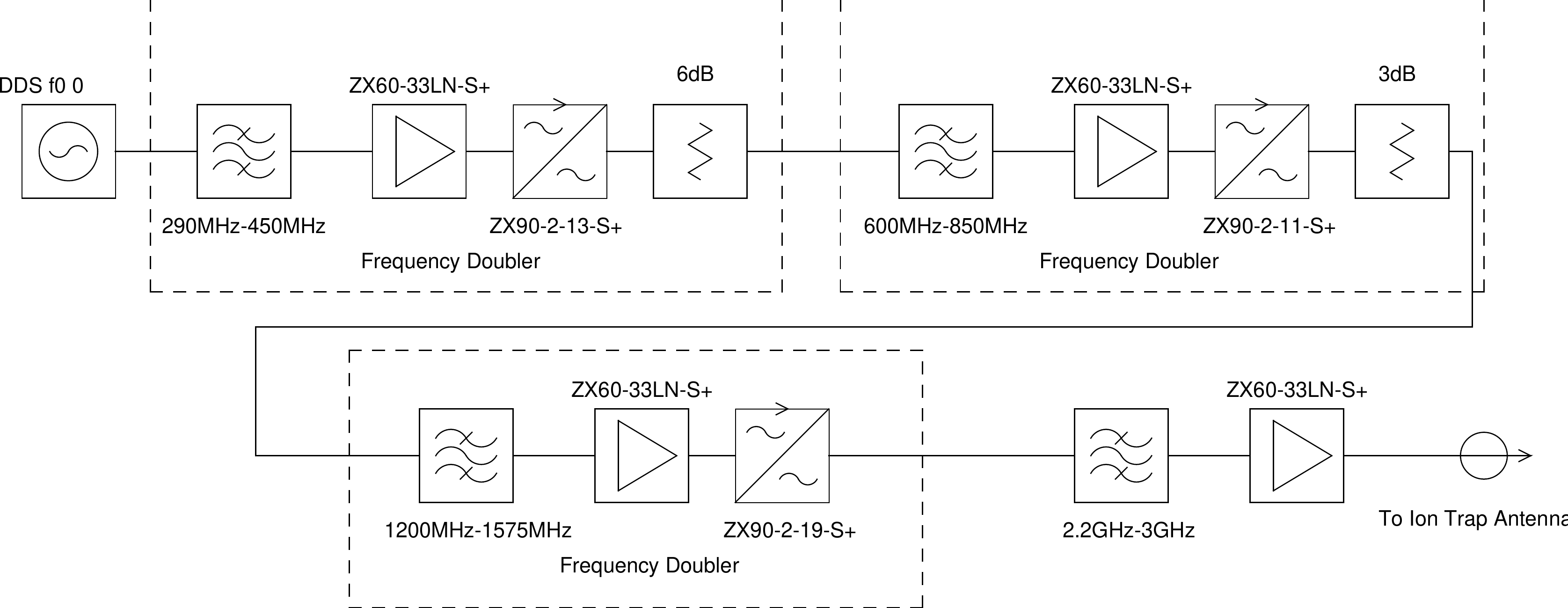}
	\caption{Schematic showing frequency multiplication from a chain of frequency doublers. Part numbers are Minicircuits.}
	\label{fig:octupler}
\end{figure*}

\begin{figure*}
	\centering
	\includegraphics[width=0.8\textwidth]{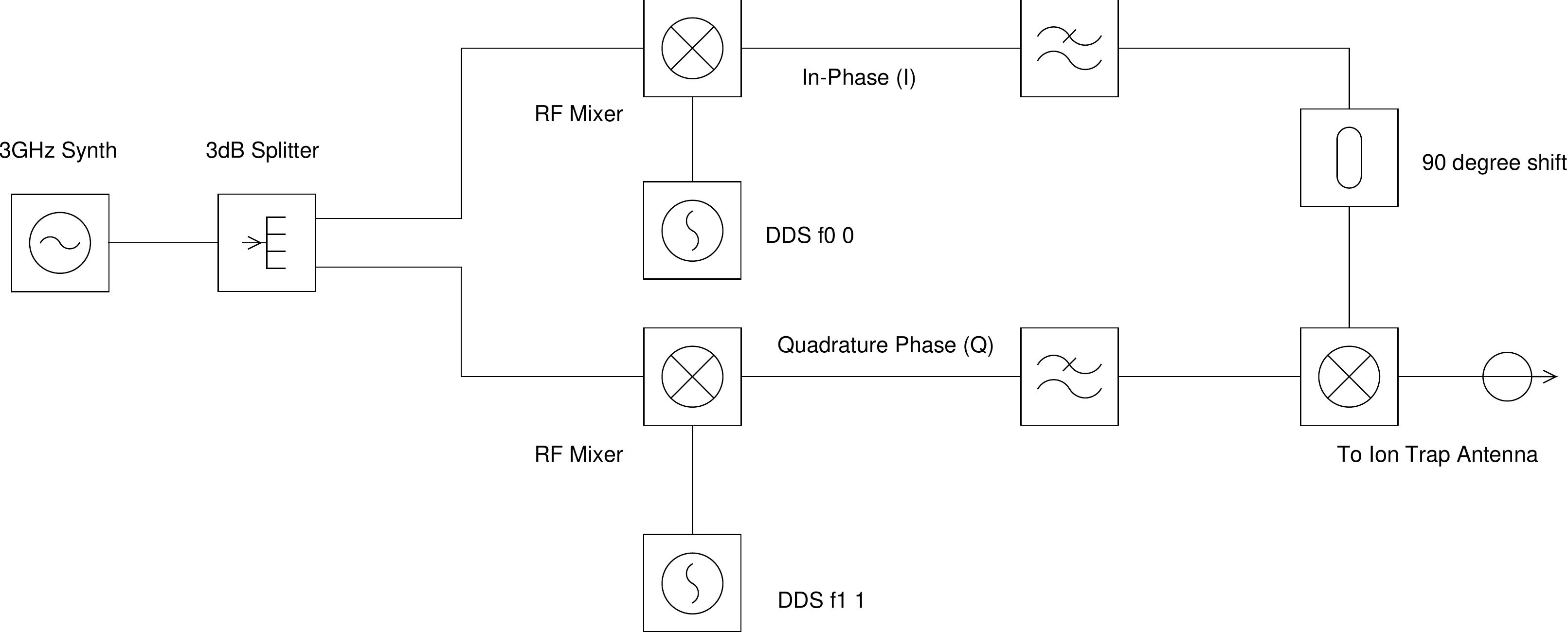}
	\caption{Hartley scheme used for single-sideband generation using two DDS channels.}
	\label{fig:hartley}
\end{figure*}

Two methods have been employed to reach the higher frequencies required for addressing hyperfine transitions.
The first is a chain of frequency couplers as shown in figure \ref{fig:octupler}.
Commercial frequency doublers use the nonlinear properties of a P-N junction to create second harmonics.
However, careful filtering and amplifying is required, giving a narrowband solution that can be sensitive to thermal drift. 

A common method used for single-side-band (SSB) transmission systems is the Hartley mixer arrangement.
This is shown in figure \ref{fig:hartley} with the slight modification of using two separate channels of the DDS system to create the in-phase and quadrature (\SI{-90}{\deg}) local oscillator signals.
This has the advantage that power and phase fluctuations can be tuned out digitally.
The hyperfine frequency is around \SI{3.2}{\giga \hertz} which is generated using an Agilent E8247C synthesiser.

\section{Conclusions}\label{sec:conclusions}

The Milldown channel card has phase noise characteristics sufficient for performing qubit gates at fault-tolerant levels, and has capabilities that will be sufficient for demanding quantum experiments now and in the near future.
It outperforms previous systems such as \cite{pruttivarasin_compact_2015} and is easily scalable by design.
The hardware was conceived and designed at the ETH Z\"urich, based on an early design from the University of Innsbruck.
They are manufactured and sold by Enterpoint Ltd.
The rf flexibility and real-time feedback capabilities have enabled novel trapped-ion experimental protocols involving dissipative state preparation, manipulation and tomography\,\cite{kienzler_quantum_2015, lo_spin-motion_2015, kienzler_observation_2016} and their application to investigating quantum foundations\,\cite{fluhmann_sequential_2017}, as well as ion transport-based quantum gates\,\cite{de_clercq_parallel_2016, de_clercq_estimation_2016}.
The low-latency feedback has been useful in quantum contextuality experiments\,\cite{leupold_sustained_2017}, and more elaborate experiments involving feedback are planned.
Additionally the system has been used in single- and two-qubit gate demonstrations with fidelities of 99.9934(3)\,\% and 99.9(1)\,\% respectively, above the fault-tolerant thresholds for many quantum error correction schemes\,\cite{harty_high-fidelity_2014, ballance_high-fidelity_2016}.
Gates using near-field microwaves\,\cite{harty_high-fidelity_2016-1} using the cards with techniques discussed in section \ref{sec:frequency_upconversion} have achieved comparable fidelities.

\section{Acknowledgements}\label{acknowledgements}

We thank Edwin Dornbierer, Daniel Kienzler, Hsiang-Yu Lo, Matteo Marinelli and David Nadlinger for significant contributions to the firmware and software discussed above, and Jonathan Home for helpful comments on the manuscript.
The rf design was partially based on work generously shared by Philipp Schindler and others from the University of Innsbruck.
We acknowledge support from the Swiss National Science Foundation under grant numbers 200021\_134776 and 200020\_153430, and through the National Centre of Competence in Research for Quantum Science and Technology (QSIT).
This work has been partially supported by the U.K. Engineering and Physical Sciences Research Council via the Quantum Technology hub for Networked Quantum Information Technologies (EP/M013243/1), as well as by the Office of the Director of National Intelligence (ODNI), Intelligence Advanced Research Projects Activity (IARPA), via the U.S. Army Research Office Grant No. W911NF-16-1-0070.

\bibliography{bib_refs}
\end{document}